\documentclass[aps,preprint,onecolumn,amsmath,amssymb,floatfix,longbibliography]{revtex4-2}

\usepackage{color}
\usepackage{mathrsfs}
\usepackage{graphicx}
\usepackage{dcolumn}
\usepackage{bm}
\usepackage{hyperref}
\usepackage{amsmath,amsfonts,amssymb,ulem}
\usepackage{epstopdf}
\usepackage{xcolor}
\usepackage{multirow}
\usepackage{newtxtext,newtxmath}

\usepackage{graphicx}

\begin{document}

\title{A high-intensity laser-based positron source}
\author{S. S. Bulanov$^1$}\email[]{sbulanov@lbl.gov}
\author{C. Benedetti$^1$}
\author{D. Terzani$^1$}
\author{C. B. Schroeder$^1$}
\author{E. Esarey$^1$}
\affiliation{$^1$Lawrence Berkeley National Laboratory, Berkeley, California 94720, USA}
\author{T. Blackburn$^2$} 
\author{M. Marklund$^2$} 
\affiliation{$^2$Department of Physics, University of Gothenburg,
SE-41296 Gothenburg, Sweden}

\begin{abstract} 
Plasma based acceleration is considered a promising concept for the next generation of linear electron-positron colliders. Despite the great progress achieved over last twenty years in laser technology, laser and beam driven particle acceleration, and special target availability, positron acceleration remains significantly underdeveloped if compared to electron acceleration. This is due to both the specifics of the plasma-based acceleration, and the lack of adequate positron sources tailored for the subsequent plasma based acceleration. Here a positron source based on the collision of a high energy electron beam with a high intensity laser pulse is proposed. The source relies on the subsequent multi-photon Compton and Breit-Wheeleer processes to generate an electron-positron pair out of a high energy photon emitted by an electron. Due  to the strong dependence of the Breit-Wheeler process rate on photon energy and field strength, positrons are created with low divergence in a small volume around the peak of the laser pulse. The resulting low emittance in the submicron range potentially makes such positron source interesting for collider applications.    
\end{abstract}

\maketitle

\section{introduction}

Recent progress in development of high power laser systems led to the construction of a number of high power laser facilities \cite{danson.hplse.2019}.
One of the main applications of these lasers is plasma based particle acceleration. Whereas the acceleration of ions is, with a couple of exceptions, mostly at the proof-of-principle stage, and the maximum ion energy only recently was reported to exceed 100 MeV (see \cite{dover.lsa.2023} and references cited therein), the acceleration of electrons and, to some extent, positrons \cite{tajima.prl.1979, esarey.rmp.2009, geddes.af6.2022} reached the level where their application in high energy physics, material science, and biology is discussed. Plasma-based acceleration of electrons was extensively studied over the last two decades, resulting in maximum electron energy going from several hundred MeV \cite{geddes.nature.2004, faure.nature.2004, mangles.nature.2004} to almost 8 GeV \cite{gonsalves.prl.2019}. High efficiency acceleration and low energy spread electron beams were demonstrated experimentally \cite{gonsalves.prl.2019, litos.nature.2014, lindstrom.prab.2021}. However, plasma-based acceleration of positrons turned out to be much more challenging \cite{cao2023positron}.

\begin{figure}
    \includegraphics[width=0.8\linewidth]{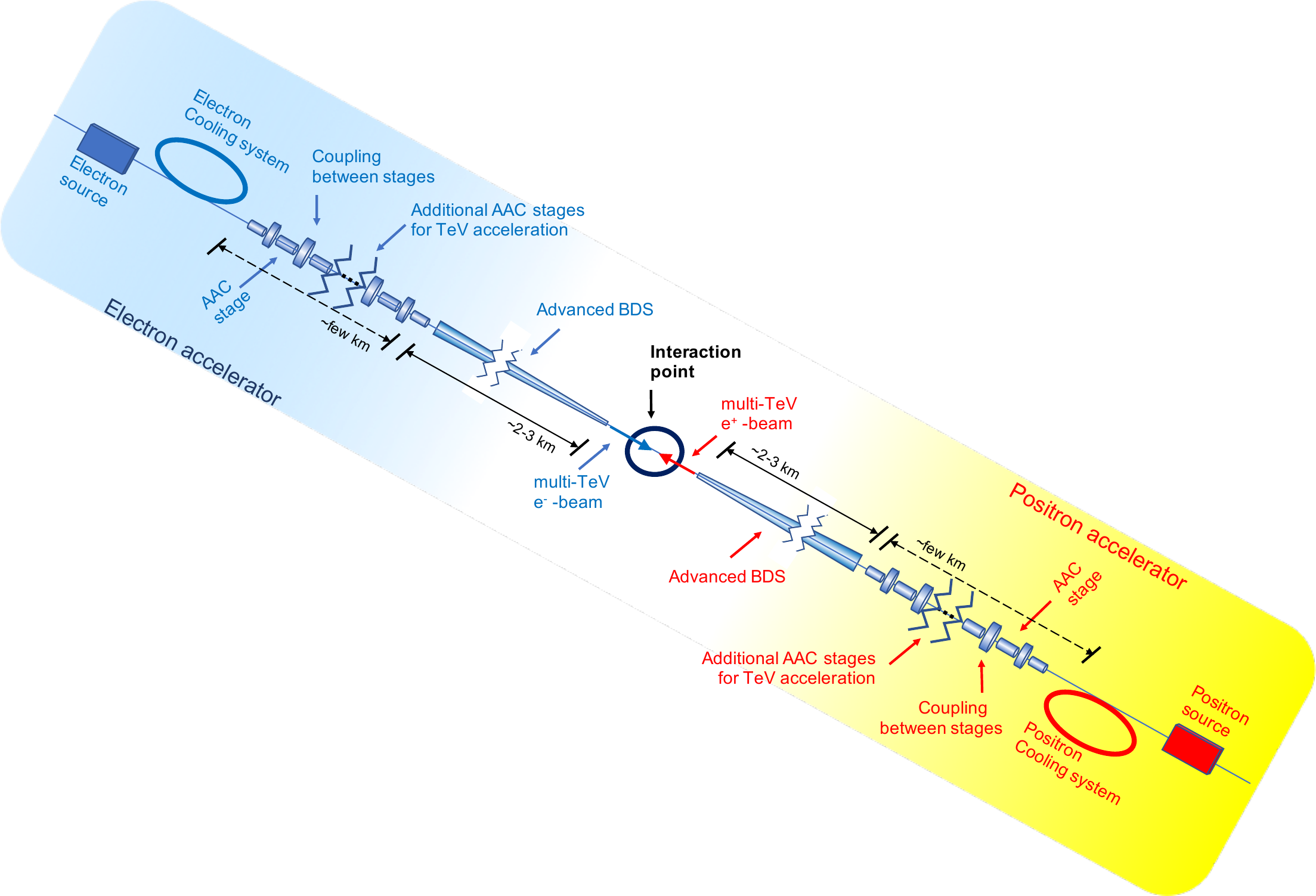}
    \caption{The principle scheme of the TeV-class plasma-based electron-positron collider. Reproduced from \cite{geddes.af6.2022}}
    \label{fig:tev_collider_scheme}
    \end{figure}

Plasma-based accelerators can have accelerating gradients of 10s to 100s GeV/m. This makes them attractive candidates for future TeV-scale linear electron-positron colliders, $\gamma\gamma$ colliders, and compact sources of high-frequency radiation \cite{roadmap.doe.2016, joshi.pop.2020, benedetti.arxiv.2022a, benedetti.arxiv.2022b, muggli.arxiv.2022}.
It is envisioned that a plasma based collider will have two symmetric arms, which would feature staged acceleration of electrons and positrons, respectively (see Fig. \ref{fig:tev_collider_scheme}). These stages will be either laser or particle beam powered (see \cite{geddes.af6.2022} for details).
The electrons and positrons will be produced in a low emittance source and, if necessary, cooled before entering the staged accelerator. An advanced beam delivery system is needed to transport the beams to the interaction point and ensure that the required luminosity goal is reached.  

%While the compact sources of high frequency radiation and $\gamma\gamma$ colliders would require only electrons to be accelerated, an electron-positron collider would need the positrons to be accelerated also. 

In conventional accelerators positrons are produced in the  collisions of high energy electron or photon beams with solid density high-Z foils (see ILC and CLIC designs \cite{seimiya.ptep.2015}).
Then, a small portion of the generated positrons is captured to be accumulated and cooled down in a damping ring for further acceleration. Plasma-based accelerators are expected to be compact by design, and this leads to a set of unique requirements on the properties of the positron source, such as, for instance, having an efficient source which produces short bunches suitable for injection into the plasma-based accelerator. 

In order to meet these requirements several positron source concepts were recently proposed.
The most straightforward concept is represented by the interaction of a high-power laser with a solid-density target which is several millimeters thick \cite{chen.prl.2009,chen.prl.2015,liang.sr.2015}.
Here, the electrons accelerated by the laser at the front surface propagate through the target emitting photons along the way due to bremsstrahlung.
These photons create electron-positron pairs in the course of their interaction with the nuclei in the target \cite{gahn.apl.2000,sarri.prl.2013,sarri.ncomm.2015,roadmap.doe.2016,alegro.arxiv.2019,alegro.arxiv.2020}.
%In principle, a high-energy electron beam can be used instead of the laser pulse in a positron production scheme, which is typical for conventional positron sources. In such scheme a high power laser is used to accelerate electrons via laser plasma acceleration \cite{gahn.apl.2000,sarri.prl.2013,sarri.ncomm.2015,roadmap.doe.2016,alegro.arxiv.2019,alegro.arxiv.2020}. 
An additional feature of this scheme is that sheath electric fields are generated at the back of the target (like in the Target Normal Sheath Acceleration \cite{daido.rpp.2012} regime of laser ion acceleration) providing a boost to the positron energy.
A high-energy electron beam can be used instead of the high-power laser pulse. In this case it was shown that, for a sufficiently intense beam, the coherent transition radiation generated at the target/vacuum interface can accelerate the positrons at the back of the target \cite{zhangli.commphys.2020}. 

\begin{figure}
    \includegraphics[width=0.8\linewidth]{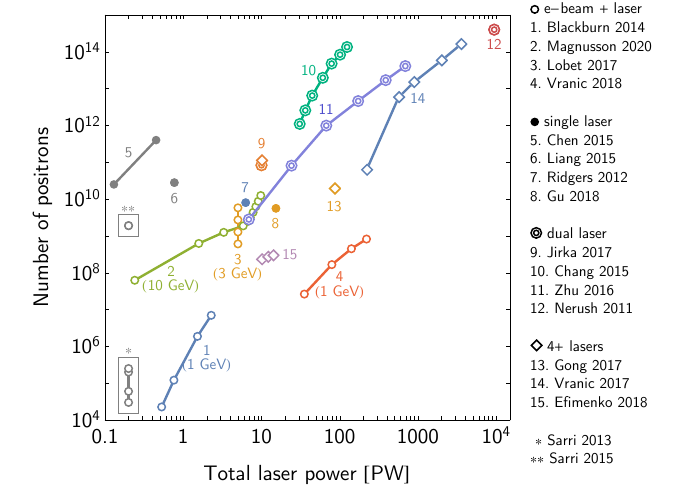}
    \caption{%
        Number of positrons produced in high-intensity laser-plasma interactions.
        For laser-electron beam interactions (open circles), the energy of the electron beam is noted in brackets.
        Points marked with asterisks indicate experimental results from LPA electron-beam interactions with high-Z foils \cite{sarri.prl.2013,sarri.ncomm.2015}; in these cases the laser power is not indicated. Reproduced from \cite{gonoskov.rmp.2022}.
    }
    \label{fig:PositronYield}
    \end{figure}

Positron production using high-intensity lasers was extensively studied over the last two decades employing a number of different mechanisms and interaction setups mostly analytically and in computer simulations \cite{dipiazza.rmp.2012, gonoskov.rmp.2022, fedotov.arxiv.2022} (see Fig. \ref{fig:PositronYield}). We note that the production of electron-positron pairs is very sensitive to the EM field strength, e.g., varying the laser intensity by four orders of magnitude results in a ten orders of magnitude variation of the number of positrons.
The most straightforward mechanism considered is a high energy electron beam interaction with a high intensity laser pulse \cite{blackburn.prl.2014, lobet.prab.2017, vranic.sr.2018, magnusson.prl.2019}.
Other mechanisms include either a laser pulse or an e-beam collision with fixed plasma targets of different density, thickness, and composition \cite{nerush.prl.2011, ridgers.prl.2012, chang.pre.2015, zhu.ncomm.2016, gong.pre.2017, jirka.sr.2017, vranic.ppcf.2017, gu.cp.2018, efimenko.sr.2018}. The summary of these studies is shown in Fig. \ref{fig:PositronYield}, where the number of positrons produced in a specific interaction setup is plotted for the total laser power used. The only experimental results shown in this figure come from an LPA electron beam interaction with high-Z foils, that is why there is no laser power assigned to them. We note that a realistic laser should be a short duration PW-class one, otherwise the application requirements in terms of repetition rate would be extremely hard to satisfy. Previous results indicate that $10^7-10^8$ positrons, i.e., $10$s of $\text{pC}$, can be expected to be produced in this case.  

Most of the positron production concepts demonstrate poor positron beam quality, which results in significant difficulties in capturing and further accelerating the positron beam in a plasma-based accelerator. The large angular divergence that the positrons acquire when they are created in the solid density foil is a challenge for the beam transport from the source to the acceleration stage (see e.g. \cite{amorim.ppcf.2023,terzani.arxiv.2023}). 

%Moreover, the number of positrons created is much smaller than that of the initial electrons, which means that high charge electron beams are needed.   

In order to address these challenges as well as introduce a positron source scheme that can be tested at existing PW-class laser facilities we propose to optimize the interaction of an energetic electron beam with a counter-propagating high intensity laser pulse. We assume that this interaction is made possible by splitting the laser energy in two parts, where one is used to accelerate electrons via LPA and the other is used to generate high-intensity electromagnetic (EM) field.  In this configuration electrons emit high energy photons in the EM field of a laser via the multi-photon Compton effect, and some part of these photons decays into electron-positron pairs under the action of the laser EM field via the multi-photon Breit-Wheeler effect. These two effects are described in the framework of strong field quantum electrodynamics (SFQED) \cite{ritus.jslr.1985, dipiazza.rmp.2012,gonoskov.rmp.2022,fedotov.arxiv.2022} and their strength is characterized by the parameters $\chi_e$ and $\chi_\gamma$,
\begin{equation}
    \chi_e=\frac{\sqrt{|p_\mu F^{\mu\nu}|^2}}{m_e E_{cr}},~~~\chi_\gamma=\frac{\hbar\sqrt{|k_\mu F^{\mu\nu}|^2}}{m_e E_{cr}},
\end{equation}
where $E_{cr}=m^2 c^3/e\hbar\simeq 1.32\times10^{18}\,\text{V/m}$ is the QED critical field, $p_\mu=\gamma_e m (c,\mathbf{v}_e)$ and $\hbar k_\mu=(\hbar\omega/c)(1,\mathbf{n})$ are electron and photon 4-momenta, respectively, $\hbar$ is the reduced Planck constant, $c$ is the speed of light, $e$ and $m_e$ are electron charge and mass, and $F^{\mu\nu}$ is the EM field tensor.
Here $\gamma_e=(1-\mathbf{v}_e^2/c^2)^{-1/2}$ is the electron Lorentz factor and $\mathbf{n}$ is the unit vector in the direction of photon propagation.
Typical values of $\chi_e$ and $\chi_\gamma$ that can be expected in the electron beam interaction with a laser pulse, given that a total laser power available is of the order of 1 PW are $\chi_{e,\gamma}\sim 2 - 4$ \cite{turner.epjd.2022}.
We note that the EM field of the laser pulse is usually characterized in terms of the frequency, $\omega$, and the normalized field strength $a_0=eE_0/m_e c\omega$ that can be expressed in function of the laser power $P$, wavelength $\lambda=2\pi c/\omega$, and waist $w_0$ as $ a_0=\lambda\sqrt{46.5\times 10^3 \left(P\left[\text{PW}\right]\right)}/w_0$ .
If $a_0\gg 1$ the interaction is nonlinear in the number of participating photons. In what follows we study the regime of interaction that is characterized by $a_0\gg 1$, $\chi_e,~\chi_\gamma\geq 1$, which justifies working in the framework of SFQED. 

The  positron source proposed here can be utilized at the now available PW-class laser facilities, such as ZEUS, BELLA, ELI BL, CORELS, and J-KAREN, to name a few (see Refs. \cite{danson.hplse.2019, gonoskov.rmp.2022} for the list of PW-class facilities available and under construction).
In view of this we only consider laser and electron beam parameters, i.e., peak intensity and maximum electron energy, which can in principle be achieved at such facilities. We assume that the laser beam can be split in two beams: (i)
to accelerate electrons via LPA, and (ii)  to generate high intensity EM field. By varying the ratio of beam splitting we optimize the photon and electron-positron pair production. Further  optimization can be achieved by tighter e-beam focusing,  due to the strong dependence of the Breit-Wheeler effect on the EM field strength. The photons interacting with the peak of the EM field are much more likely to produce an electron-positron pair. The produced positron beams have a low angular divergence and emittance, which makes such a source an interesting option to consider for implementation in a future plasma-based lepton collider.   
    
The paper is organized as follows. First, we discuss the principle scheme of the positron source in section \ref{sec:positron_source}. Then, in Section \ref{sec:typical_interaction}, we present the results of numerical simulations of a high-energy electron beam interaction with a high-intensity laser pulse. We discuss the dependence of the positron production effectiveness on total laser energy and the ratio of beam splitting in section \ref{sec:positron_charge}. We show the optimization of the positron source via employing tighter focused electron beams in section \ref{sec:positron_source optimization}. In section \ref{sec:positron_source emittance} we demonstrate that the optimized positron source can produce beams with extremely low emittance. We conclude in Section \ref{sec:conclusions}.  

\section{Two beamline PW-class laser facility as a positron source}\label{sec:positron_source}

\begin{figure}
    \includegraphics[width=0.8\linewidth]{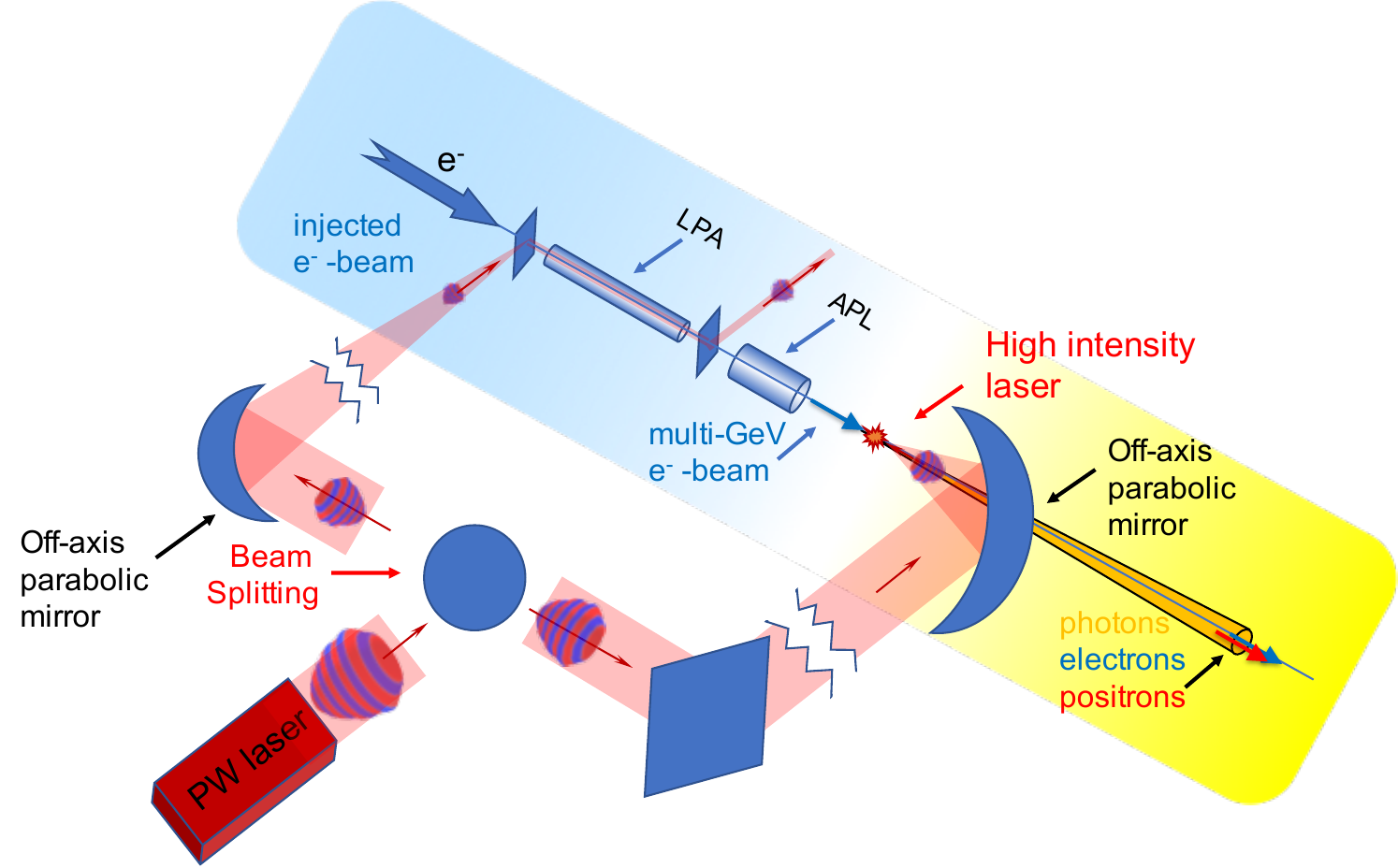}
    \caption{The principal scheme of the high laser intensity positron source based on the laser - e-beam collision. The source features a single PW-class laser, which is split into two beams to power stage 1 (LPA of electrons) and stage 2 (high intensity pulse). The electrons coming out of the LPA are focused by the APL at the IP, where they collide with a high intensity laser pulse.}
    \label{fig:scheme}
    \end{figure}

The proposed positron source is based on two consecutive SFQED processes: multi-photon Compton (MC) and multi-photon Breit-Wheeler (MBW), where the former is responsible for the photon emission and the latter for the photon decay into an electron-positron pair, both happening in a strong EM field.
This is realized by colliding a high-energy electron beam with a high-intensity laser pulse. As it is envisioned that this positron source is an "all-optical" one \cite{bulanov.nima.2011}, i.e., the electron beam is accelerated by a laser in a plasma, the following two-stage setup is proposed (see Fig. \ref{fig:scheme}). The positron source uses a single PW-class laser which is split into two beams powering the two stages of the setup. The first stage is an LPA featuring a long focal length parabola needed to deliver a laser pulse with $I\sim 10^{18}$ W/cm$^2$ intensity to an underdense plasma target.
Quasi-monoenergetic electron beams with multi-GeV energy and low divergence can be obtained using a PW-class laser (see, e.g., \cite{gonsalves.prl.2019}). The high-energy electron beam then goes through an Active Plasma Lens (APL) \cite{tilborg.prl.2015, tilborg.prab.2017, pompili.prl.2018} to be focused at the interaction point (IP).
The APL focuses a specific part of the electron beam with certain energy and energy spread determined by the APL parameters.
Here we assumed that the electron beam can be modeled by a Gaussian distribution with rms dimensions ($\sigma_x$, $\sigma_y$, $\sigma_z$), energy spread $\Delta\mathcal{E}$, and divergence $\theta$.
The second stage delivers a high-intensity (i.e., $I \sim 10^{\mathord{21-22}}\,\text{W/cm}^2$) laser pulse to the IP by employing a short focal length parabola.
In what follows we assume a head-on collision of the electron beam and the laser pulse.
In this case the short focal length parabola can have a hole to allow for the passage of the electron beam and secondary particles, i.e., photons and electron-positron pairs generated during the collision.
In principle, collision at an angle is also possible and will be considered in a future publication.    

Here we do not address the question of the subsequent positron beam capture and acceleration in the plasma based stage. We characterize the beam by average energy, energy spread, angular divergence, and emittance, and compare them to the one obtained in the e-beam foil interaction schemes, where the capture and acceleration were demonstrated in numerical simulations \cite{amorim.ppcf.2023,terzani.arxiv.2023}.    

\section{Typical e-beam laser interaction}\label{sec:typical_interaction}

In order to illustrate the interaction of a high energy e-beam with a high intensity laser we show simulation results of a 9.1 GeV e-beam with a $a_0=40$ laser pulse. This corresponds to a 25 J /15 J energy split between stages one and two (see Appendix). The electron beam was chosen to have a 3\% energy spread, 2 $\mu$m length, and $r_b=2.5$ $\mu$m radius. The divergence was 0.52 mrad. The electrons were normally distributed. The laser pulse was focused down to a spot $w_0=2.5$ $\mu$m and had a 10 cycle duration, $\lambda=800$ nm. The electron beam radius was chosen to be equal to the laser waist at focus to maximize the interaction volume. (We will address the dependence of the interaction volume on relative transverse dimensions of the  electron beam and the laser below.)
The simulations were performed using a single-particle code {\sc ptarmigan} \cite{blackburn.njp.2021,blackburn.epjc.2022}, which takes into account SFQED effects, namely MPC and MBW, accounting for the angular distribution of the secondary particles.
The code can employ both local constant field approximation (LCFA) and local monochromatic wave approximation (LMA) to  calculate the Compton and Breit-Wheeler rates, as well as treat the interaction in terms of classical Landau and Lifshits equation of motion.
This is in contrast to all the particle-in-cell (PIC) codes with QED modules (see \cite{gonoskov.rmp.2022} and references cited therein for PIC QED codes descriptions), where the approximation of the collinear emission is used.
This approximation states that the photons and electron-positron pairs generated in MC and MBW processes are assumed to have their momenta directed along the instantaneous momentum of an initial state particle. Since the code {\it ptarmigan} is a single-particle code, collective effects such as space-charge are not modeled, but they are expected to be negligible for the parameters considered in this study. 

\begin{figure}[tb]
\includegraphics[width=0.95\columnwidth]{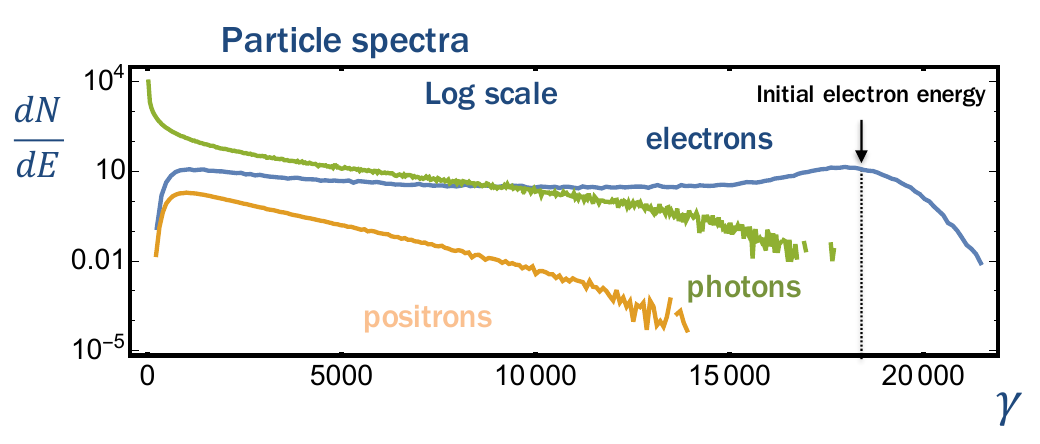}
\caption{The final spectra, on a logarithmic scale, of electrons, photons, and positrons after the collision of a 9.1 GeV electron beam with an $a_0=40$ laser pulse.} \label{fig:spectra}
\end{figure}

%As it was mentioned above, the interaction of charged particles and photons with strong EM fields is characterized by two invariant parameters, $\chi_e$ and $\chi_\gamma$.
For the parameters chosen for the simulation $\chi_e$ and $\chi_\gamma$ can acquire values up to 4, which indicates that the interaction is deep in the quantum regime of SFQED.
Moreover, for $a_0\approx 40$ and $\mathcal{E}_e\approx 9$ GeV the mean free path of an electron with respect to radiating a photon is about $\sim\lambda/5$.
Since an electron on average loses a fraction (16/63) of its energy on radiation during the MC process \cite{ritus.jslr.1985, jirka.pra.2021}, one can expect multiple photon emissions by each electron of the beam, as well as a significant energy loss.
As for the photons, those with energy higher than $\hbar\omega_{th}$, which is defined as $L_{rad}(\hbar\omega_{th})=c\tau$, will predominantly decay into electron-positron pairs via the MBW process, where $\tau$ is the laser pulse duration.
For the parameters of the electron beam and the high-intensity laser chosen for our simulations the estimate gives several percent of photons decay in electron-positron pairs.
The results of the simulations agree with these estimates.
The electron beam loses 60\% of its initial energy during the interaction, and develops an almost 100\% relative energy spread (see Fig. \ref{fig:spectra}). 
%%%%

The electron beam energy was mainly transformed into the energy of photons, with each electron emitting 20 photons on average. A small part of these photons was converted into $e^+e^-$ pairs via the multi-photon Breit-Wheeler process, amounting to one pair per 200 initial electrons. The final spectra of electrons, positrons, and photons are shown in Fig. \ref{fig:spectra}.  

%As it was mentioned above the head-on collision of the e-beam and the laser pulse is attractive for positron production because of the narrow angular spread of secondary particles. The photons are predominantly emitted into the $1/\gamma$ angle, however, the transverse motion of electrons in the laser pulse brings in an additional factor of $a_0$ in the laser polarization plane. Thus, it is plausible to expect the divergences at $1/\gamma$, $a_0/\gamma$ levels.

\begin{figure}[tb]
\includegraphics[width=0.95\columnwidth]{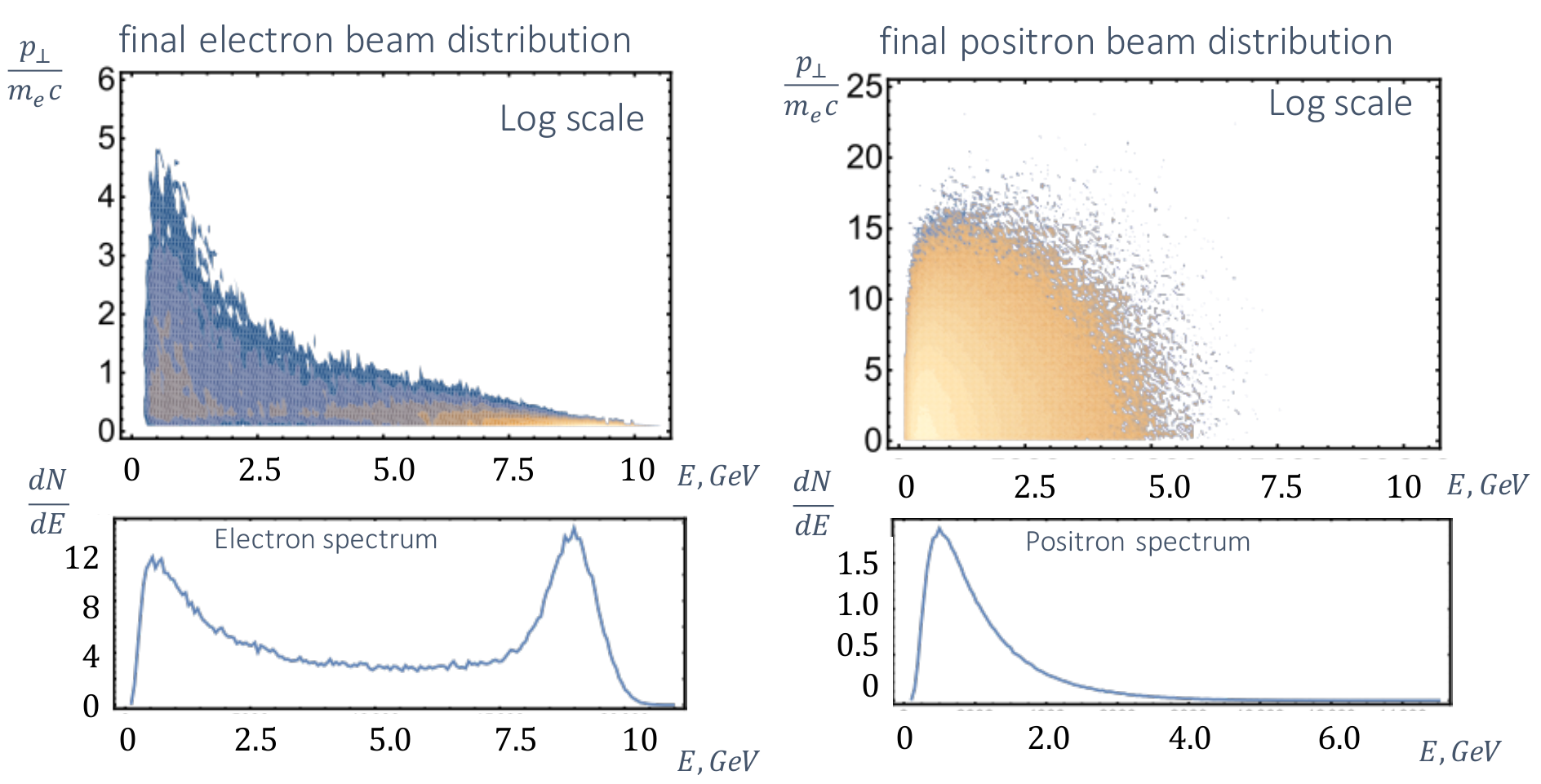}
\caption{The final electron and positron distributions in the $(\gamma,p_\perp)$ plane (a,b). The final spectra of electrons and positrons on a linear scale (c,d).} \label{fig:beam_distr}
\end{figure} 

In Fig. \ref{fig:beam_distr}, we show the final electrons and positron beam distributions in the $(\gamma,p_\perp/m_ec)$ plane.
The electron distribution stretches from the initial energy all the way back to low energies with the transverse momentum component increasing as the overall electron energy goes down \cite{blackburn.pra.2020}.
This is due to either the fact that transverse momentum is being accumulated through multiple emissions, or to the fact that the emission of a high energy photons results in a strong transverse kick to the emitting electron.
The positrons are more evenly distributed, which is due to the properties of the BW rate.
Each of the distributions figures is accompanied by a spectrum figure in linear scale to show at what energy the majority of particles reside. While positrons have a single maximum at 400 MeV, the electron spectrum shows two maxima. One corresponds to the initial energy of the electron beam, the other is governed by the strong energy dissipation due to the photon emission.

Even though $\chi_{max}=3.7$, which indicates that a quantum description of the interaction is required, one can use classical equations of motion to estimate the behavior of the low energy part of the electron population. If we consider electrons with energy around 400 MeV, their effective $\chi$ is at least 20 times smaller. We can write a 1D equation of motion for this population assuming that the energy loss is dominated by the radiation emission and that the effects of the Lorentz force are negligible. The solution of such an equation of motion is
\begin{equation}
    p_f=\frac{p_0}{1+\epsilon_{rad}\omega_0(p_0/m_e)\int_0^t a^2(-\eta)d\eta},
\end{equation}
where $\epsilon_{rad}=4\pi r_e/3\lambda$, $r_e=2.8\times 10^{-13}$ cm is the classical electron radius. We note that for $\epsilon_{rad}\omega_0(p_0/m_e)\int_0^t a^2(-\eta)d\eta\gg 1$ the final momentum does not depend on the initial one. In a dynamical systems with high dissipation of energy, particles ``forget'' about the initial condition. In this case, $p_f\approx 500$ MeV, which is very close to the value obtained in the simulations for the low energy peak in the electron spectrum. 

Thus, we observed the formation of a positron beam with central energy of 400 MeV and small angular divergence as a result of a multi-GeV electron beam collision with a high-intensity laser pulse. In order to determine whether such an interaction offers a promising source of positrons for high energy physics applications, in what follows we characterize the positron beam in terms of particle number, divergence, and emittance.  

\section{Positron beam charge}\label{sec:positron_charge}

One of the key properties of the positron source is the number of produced positrons per bunch.
For example, a sufficiently large positron beam charge removes the need for an accumulating ring and simplifies capture and subsequent acceleration of positrons in plasma-based accelerator stages. Since the number of positrons depends on the effectiveness of the SFQED processes, the parameters of the interaction should be optimized to maximize $\chi_e$ and $\chi_\gamma$.
Here we study the dependence of the positron beam charge on the ratio of the two laser beam energies, keeping other parameters, i.e., focal spot, spatial dimensions of the electron beam, etc., fixed.
Since the electron energy gain in an LPA scales as $U_1^{2/3}$, the maximum value of $\chi_e$ is proportional to $U_1^{2/3}\sqrt{1-U_1}$.
This leads to the conclusion that $\chi_e$ is maximized for $U_1\approx 4U_0/7$. In what follows we use the code {\sc ptarmigan} to model the electron beam laser collisions, where electron beam energy and laser pulse intensity scale according to the laser energy split. 

\begin{figure}[tb]
\includegraphics[width=0.95\columnwidth]{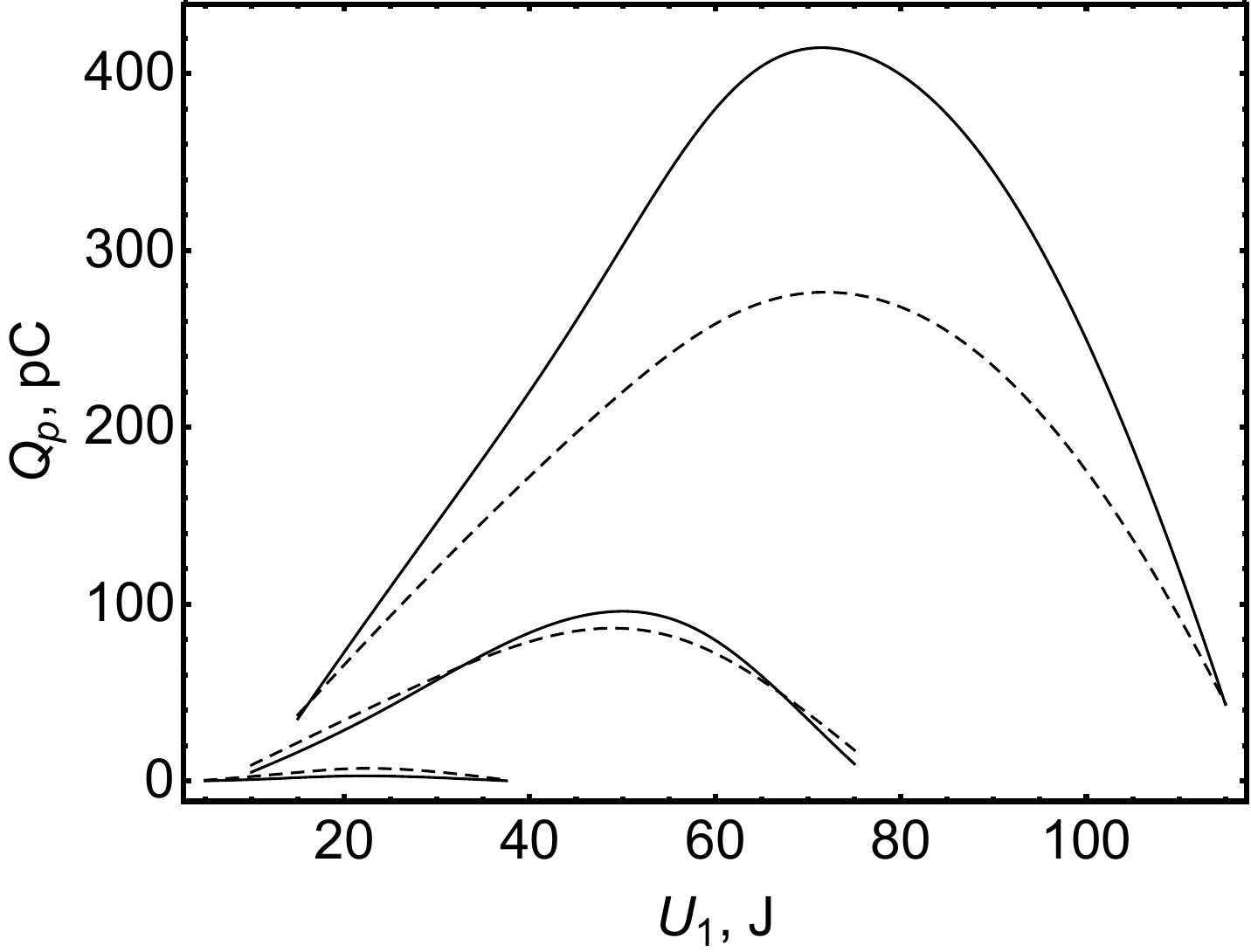}
\caption{Positron beam charge produced in the electron beam collision with a laser pulse for 40 J, 80 J, and 120 J of total laser energy. Two cases of LPA were considered for electron acceleration: externally guided or quasi-linear regime (dashed curves),  and self-guided or bubble regime (solid curves).} \label{fig:charge}
\end{figure}

We considered three cases of total laser energy used: 40, 80, and 120 J in order to illustrate the strong dependence of the positron source effectiveness on the total input energy.
The electron beam energy and charge are determined from the LPA scalings for idealized stages (see Appendix).
Two cases of LPA were considered, externally guided (operating in the quasi-linear regime) and self-guided (operating in the nonlinear bubble regime).
Since quasi-linear and bubble regimes of LPA operate at different plasma densities the total charge of an accelerated electron beam is different.
We take charge differences into account when calculating the total positron beam charge. The results of these simulations are summarized in Fig. \ref{fig:charge}, where the positron beam charge dependence on $U_1$ is shown.
All six curves demonstrate similar behavior with the maximum achieved around $U_1=U_0/2$, and significant reduction of positron beam charge for $U_1\rightarrow U_0$ or $U_1\rightarrow 0$.
We note that the quasi-linear case produces more positrons for 40 J, indicating that from the point of view of positron production, and for a relatively low total laser energy, maximizing the electron energy is more beneficial than maximizing the number of electrons in the beam.
However, as the total laser energy is increased, maximizing the number of electrons in the beam becomes the relevant factor in terms of maximizing the total positron beam charge.
In the 40 J case a positron beam with a charge as high as 7 pC was observed in simulations. This was for a 9.1 GeV, 150 pC electron beam accelerated in a quasi-linear LPA stage.
The conversion efficiency from the electron beam to positron beam is 5\% in terms of particle number and 2\% in terms of particle energy.
The utilization of higher laser energies, e.g., 80 J and 120 J leads to a significant increase in the produced positron charge, resulting in positron beams with a charge of 100s of pC.
We note that  this is partly due to the fact that higher energy laser pulses produce higher charge and energy electron beams via LPA.
However, the main contribution to the positron charge increase comes from more efficient MC and MBW processes leading to 30\% efficiency in terms of particle number for a 120 J of total laser energy.

In terms of collider applications, the number of positrons needed per bunch is around $\gtrsim$0.1 nC \cite{benedetti.arxiv.2022a,benedetti.arxiv.2022b}, which means that, for the chosen parameters of the interaction, the required laser energy is $\gtrsim 100$ J. 

We note here that the electron beam transverse size was chosen to be equal to the laser waist. While it might be a good choice to maximize the interaction volume, it is not so from the point of view of maximizing the positron production. In what follows we study the dependence of the efficiency of the positron production  on the transverse size of the electron beam. Since both Compton and Breit-Wheeler processes are intensity-dependent, there should be a strong dependence of the number of produced positrons on the transverse size of the electron beam.  

\section{Optimizing the positron source}
\label{sec:positron_source optimization}

\begin{figure}[tb]
\includegraphics[width=0.95\columnwidth]{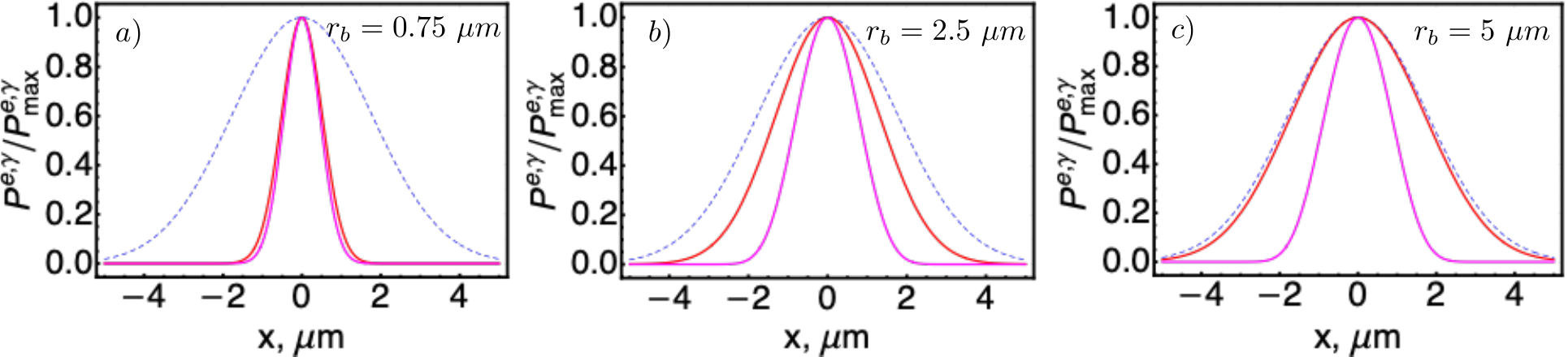}
\caption{The dependence of the MC (red curves) and MBW (magenta curves) process probabilities on the transverse profile of the laser EM field for different initial radii of the electron beam, $r_b=0.75,~2.5,~5$ $\mu$m. The probabilities are normalized to their maximum values at $x=0$. The laser EM field transverse profile is Gaussian with the width of $w_0=2.5$ $\mu$m, shown by the dashed blue curve.}
\label{fig:BW}
\end{figure}

It is well known that the $e^+e^-$ pair production by a photon in a strong EM field is exponentially suppressed at low values of $\chi_\gamma$ ($\chi_\gamma\ll 1$), in contrast to the photon emission by an electron or positron, as can be seen from the probabilities, $P^{\rm MC}$ and $P^{\rm MBW}$, of these processes \cite{ritus.jslr.1985,dipiazza.rmp.2012, gonoskov.rmp.2022,fedotov.arxiv.2022}:
The probability for the MC process is 
\begin{equation}
    P^{\rm MC}(\chi_e) =
        -\frac{\alpha c}{2\sqrt{3}\pi^2 \lambdabar_C}\frac{\chi_e}{\gamma}\int\limits_0^\infty dy\frac{5+7\zeta+5\zeta^2}{(1+\zeta)^3}K_{2/3}(y),
   \label{eq:Pmc} 
   \end{equation}
where $\zeta=(3/2)\chi_e y$, $\lambdabar_C$ is the electron Compton wavelength, $\alpha$ is the fine structure constant, and $K_{2/3}(y)$ is the Bessel function of the second kind. For $\chi_e\ll 1 $ and $\chi_e\gg 1$ the integration of Eq.~\eqref{eq:Pmc} can be carried out, and we obtain
    \begin{equation}
    P^{\rm MC}\simeq
    \begin{cases}
    0.92\frac{\alpha}{\lambdabar_C/c}\sqrt{\frac{I}{I_S}}\left(1-0.92\chi_e+...\right),~~~\chi_e\ll 1 \\ 
    0.93\frac{\alpha}{\lambdabar_C/c}\sqrt{\frac{I}{I_S}}\chi_e^{-1/3}\left(1-0.58\chi_e^{-2/3}\right),~~~\chi_e\gg 1.
    \end{cases}
    \end{equation}
The corresponding probability for the MBW process is
    \begin{equation}\label{gamma probability}
    P^{\rm MBW}=-\frac{\alpha c \chi_\gamma}{32\sqrt{3}\pi^2 \lambdabar_C}\frac{m_e c^2}{\hbar \omega}\int\limits_{\frac{8}{3\chi_\gamma}}^\infty \frac{(8\zeta+1)K_{2/3}(y)}{\zeta\sqrt{\zeta(\zeta-1)}}dy,
    \end{equation}
where $\zeta=(3/8)\chi_\gamma y$. For $\chi_\gamma\ll 1 $ and $\chi_\gamma\gg 1$ the integration of Eq.~\eqref{gamma probability} can be carried out, and we have
    \begin{equation}\label{gamma probability limits}
    P^{\rm MBW}=\begin{cases}
    0.073\frac{\alpha}{\lambdabar_C/c}\left(\frac{I}{I_S}\right)^{1/2}\exp\left(-\frac{8}{3\chi_\gamma}\right),~~~\chi_\gamma\ll 1, \\
    0.67\frac{\alpha}{\lambdabar_C/c}\left(\frac{I}{I_S}\right)^{1/2}\chi_\gamma^{-1/3},~~~\chi_\gamma \gg 1.
    \end{cases}
    \end{equation}
Thus, for a given laser pulse and a colliding electron beam with given transverse profiles, the probability of the MC process will follow either the shape of the laser or that of an electron beam. Whereas the probability of the MBW process is suppressed by the additional exponential factor, $\exp(-8/3\chi_\gamma)$. In order to illustrate this, we show in Fig. \ref{fig:BW} the dependencies of the probabilities of MC and MBW processes multiplied by the beam profile distribution, $\exp(-x^2/r_b^2)$, on the transverse coordinate $x$ for $r_b=0.75,~2.5,~5$ $\mu$m. The laser pulse width is $w_0=2.5$ $\mu$m. In the case of $r_b\ll w_0$ both MC and MBW processes are confined by the initial beam radius and almost coincide. When $r_b=w_0$, the profile of the MC process probability starts to tend towards that of the laser pulse. And, for $r_b>w_0$, the profile of the MC process probability almost coincides with that of the laser pulse, while the MBW is limited by the exponential suppression. From these results one can see that the majority of $e^+e^-$ pairs are produced near the field maximum even for a wide electron beam. So, in order to maximize the $e^+e^-$ production for a given laser pulse and electron beam, a tighter focusing for the electron beam should be employed. 

\begin{figure}[tb]
\includegraphics[width=0.95\columnwidth]{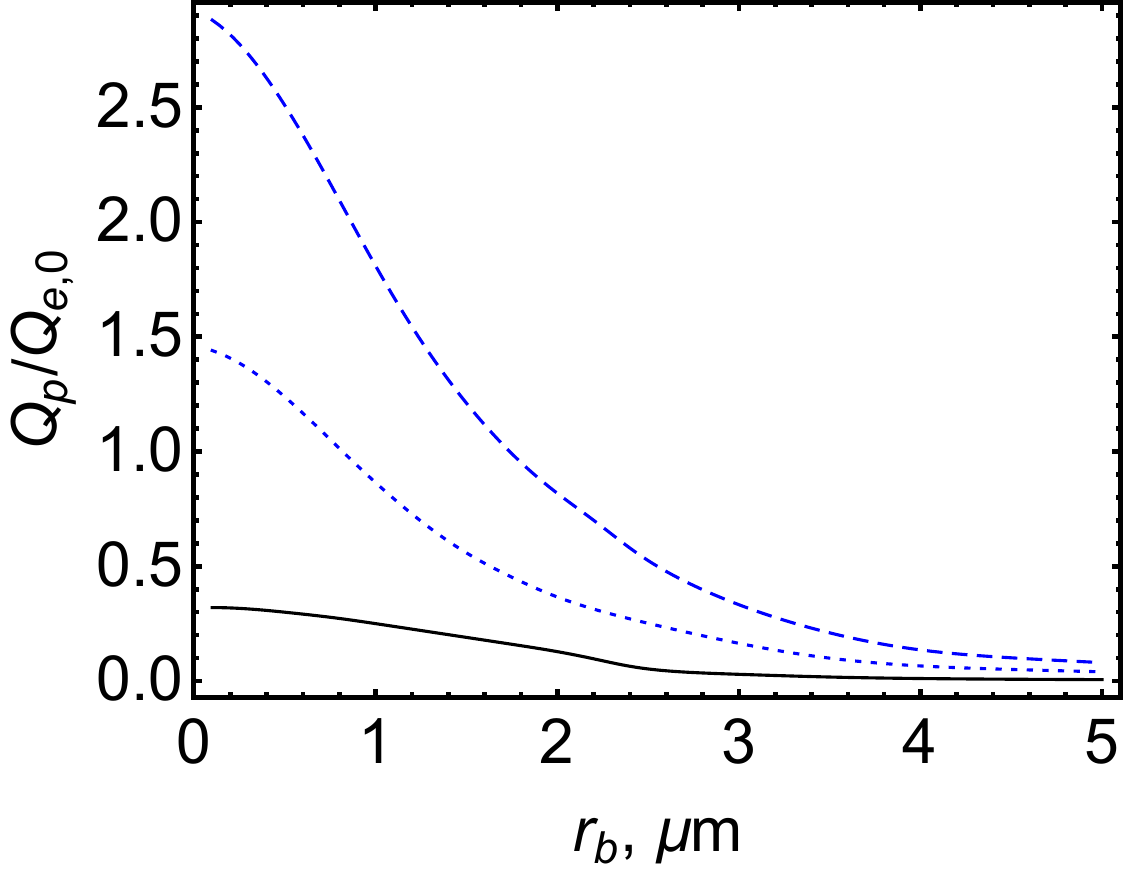}
\caption{The dependence of the positron beam charge on the initial radius of the electron beam for 40 J (solid curve), 80 J (dotted curve), and 120 J (dashed curve) total laser energy. Here $U_1=U_2=U/2$, laser focal spot radius is kept constant at $2.5~\mu$m. Electron beam parameters are 9.1 GeV energy, 2 $\mu$m length, $2.5~\mu$m radius, and 0.5 mrad divergence angle.} \label{fig:radius}
\end{figure} 

In what follows we studied the dependence of the number of produced pairs on the radius of the electron beam keeping the laser pulse waist the same, $w_0=2.5$ $\mu$m. The results of these simulations are shown in Fig. \ref{fig:radius}, where the dependence of the relative positron beam charge ($Q_p/Q_0$) on the electron beam radius is shown. As in the previous section we considered here three values of total laser energy, 40 J, 80 J, and 120 J. The electron beam radius was varied from 0.1 $\mu$m to 5 $\mu$m, while the divergence was kept constant at 0.5 mrad. In all three cases the number of produced positrons has the same dependence on the electron beam radius. The smaller is the radius, the larger is the number of positrons. And while for 40 J case the maximum number of produced positrons is 0.3 per one initial electron (or  $\sim 40$~pC), at 80 J it is 5 times larger, resulting in 1.5 positrons per initial electron (or 0.46~nC). Adding 40 J more of laser energy almost doubles this number up to 2.88 positrons per one initial electron (or 1.4 nC). Thus, one can see that the electron beam radius is a crucial parameter for the optimization of the positron source not only in terms of the positron beam charge but also in terms of lowering the beam emittance, as we will show in the next section.

\section{Production of low emittance positron beams.}
\label{sec:positron_source emittance}

Since we were motivating the study of the positron source by possible collider applications, the low emittance of produced positron beams is absolutely necessary. Typical emittances of positron sources for future colliders have submicron values \cite{chaikovska.jinst.2022,musumeci.arxiv.2022}. So here we show that similar parameters can be achieved in high-energy electron beam interaction with a high intensity laser pulse. 

The electron beam energy is chosen to be 9.1 GeV and we vary the beam radius, $r_b=0.1 - 2.5$ $\mu$m, and divergence, $\theta_e=0.1,~0.25,~0.5$ mrad. As for the high intensity laser pulse, two values of $a_0$ were considered, $a_0=20$ and $a_0=40$, to illustrate the difference between positron production at percent and tens of percent levels of the initial number of electrons. 

\begin{figure}[tb]
\includegraphics[width=0.95\columnwidth]{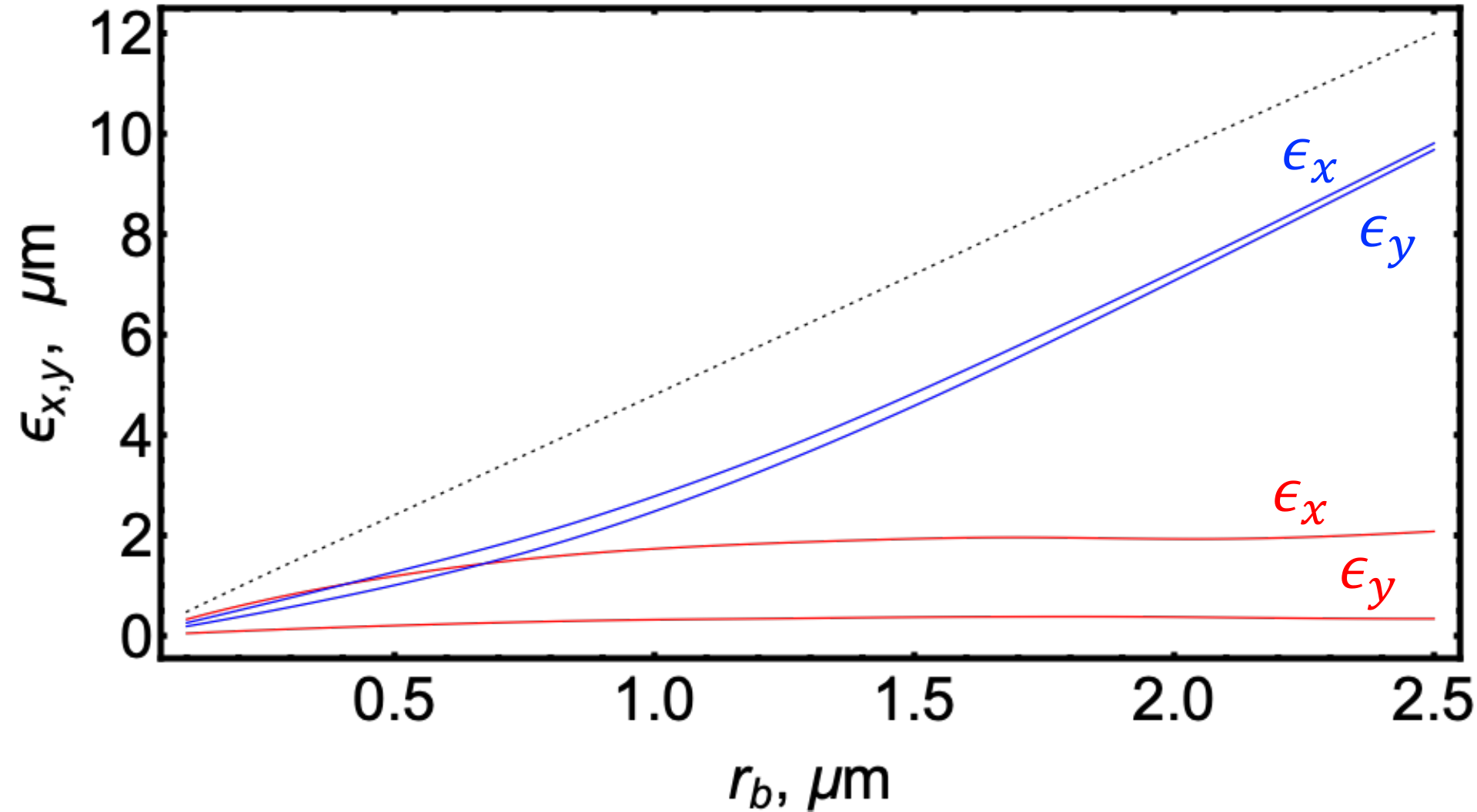}
\caption{The dependence of the normalized electron (blue curves) and positron (red curves) beam emittances, $\epsilon_x$ [$\mu$m] and $\epsilon_y$[$\mu$m] for $a_0=20$ on the initial electron beam radius. The initial electron beam normalized emittance is shown by the dotted curve.} \label{fig:emittance_a0_20}
\end{figure}

\begin{figure}[tb]
\includegraphics[width=0.95\columnwidth]{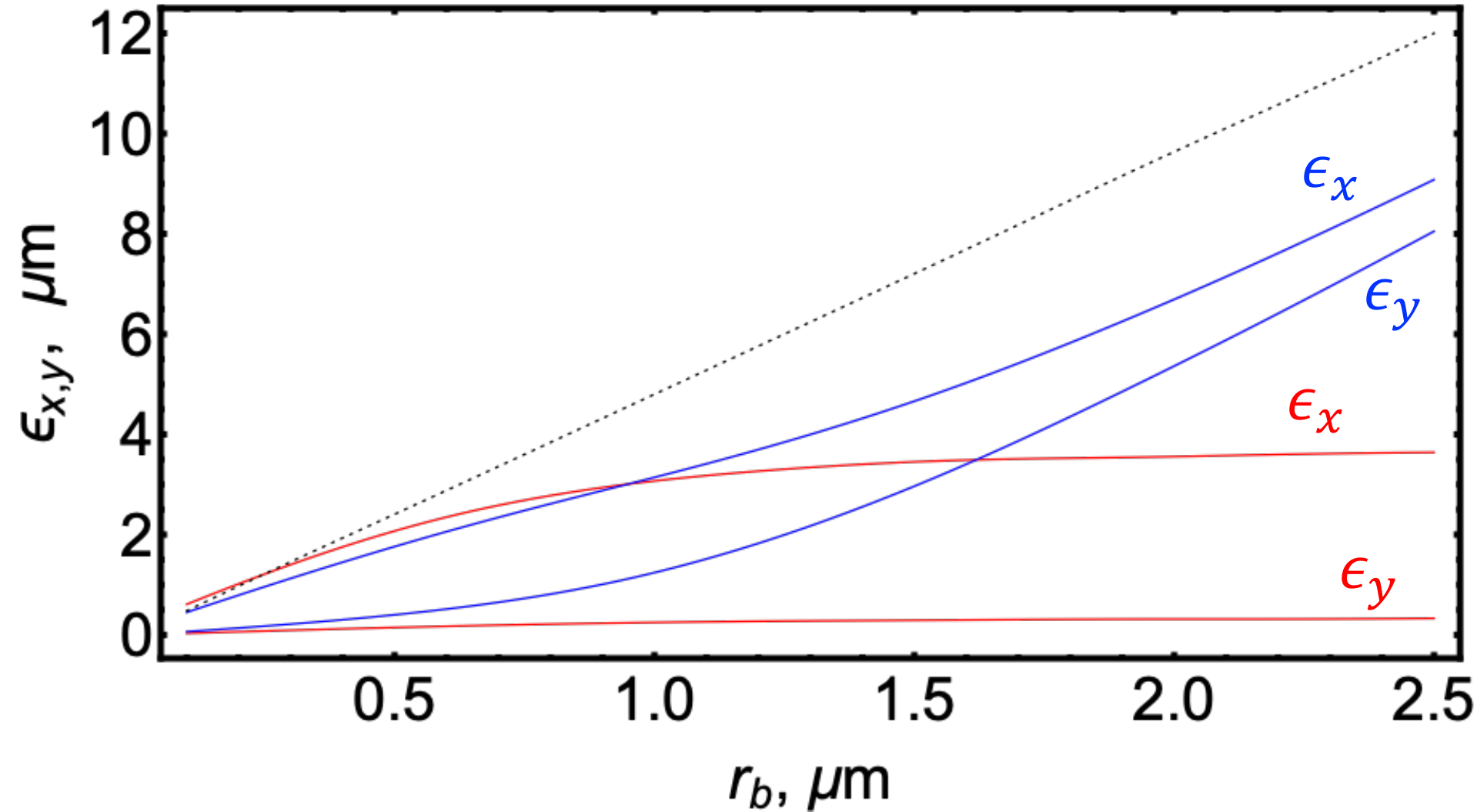}
\caption{The dependence of the normalized electron (blue curves) and positron (red curves) beam emittances, $\epsilon_x$ [$\mu$m] and $\epsilon_y$[$\mu$m] for $a_0=40$ on the initial electron beam radius. The initial electron beam normalized emittance is shown by the dotted curve.} \label{fig:emittance_a0_40}
\end{figure}

First, we address the emittance dependence on the initial radius of the electron beam. The results of {\sc ptarmigan} simulations show that the normalized transverse emittance of the positron beam increases with the increase of the electron beam radius. However, they reach values significantly lower than that of the initial electron beam. In Figs. \ref{fig:emittance_a0_20} and \ref{fig:emittance_a0_40} we show the dependencies of $\epsilon_x$ (here x is the laser polarization axis) and $\epsilon_y$ on $r_b$ for $a_0=20$ and $a_0=40$, respectively, along with the initial electron beam emittance.

The submicron emittances can be achieved for sufficiently small electron beam radii. This is in addition to the fact that small radius electron beams significantly enhance positron production for fixed laser intensity and electron beam energy, as it was shown in the previous section. In what follows we investigate if further reduction of the positron beam emittance can be achieved by reducing the emittance of the initial electron beam.  

We choose the following parameters of the electron beam: radius of $r_b=0.1$ $\mu$m and divergence of $0.1$ mrad, $0.25$ mrad, and $0.5$ mrad, which corresponds to an initial emittance of $\epsilon_x=\epsilon_y=0.093,~0.23$, and $0.48$ $\mu$m, respectively. The interaction is modeled for two values of laser intensity corresponding to $a_0=20$ and $40$. Other laser parameters are the same as in the simulations above. The results of the simulations are summarized in the Table \ref{tab:1omega}.

\begin{table}[]
\begin{tabular}{|c||cc|cc|}
\hline
\multirow{2}{*}{$\epsilon_x/\epsilon_y$} & \multicolumn{2}{c|}{$a_0=20,~\omega=\omega_0$}    & \multicolumn{2}{c|}{$a_0=40,~\omega=\omega_0$}    \\ \cline{2-5} 
                  & \multicolumn{1}{l|}{electron} & positron & \multicolumn{1}{l|}{electron} & positron \\ \hline \hline
 $\theta=0.1,~\epsilon_0=0.1$ $\mu$m     & \multicolumn{1}{l|}{0.18/0.038} & 0.32/0.025 & \multicolumn{1}{l|}{0.44/0.02} & 0.61/0.02 \\ \hline
 $\theta=0.25,~\epsilon_0=0.2$    & \multicolumn{1}{l|}{0.2/0.089} & 0.33/0.05 & \multicolumn{1}{l|}{0.46/0.03} & 0.61/0.03 \\ \hline
 $\theta=0.5,~\epsilon_0=0.48$     & \multicolumn{1}{l|}{0.26/0.2} & 0.34/0.1 & \multicolumn{1}{l|}{0.45/0.07} & 0.62/0.06 \\ \hline
\end{tabular}
\caption{The normalized emittances, $\epsilon_x$ and $\epsilon_y$ in $\mu$m, of electron and positron beams after the interaction with either $a_0=20$ or $a_0=40$ laser pulse for three initial values of electron beam divergence $\theta=0.1,~0.25,~0.5$ mrad. For $a_0=40$, all three cases of initial divergence result in the production of a 40 pC positron beam. For $a_0=20$ the positron beam charge is down to 2.4 pC.}
    \label{tab:1omega}
\end{table}

Both electron and positron emittances demonstrate similar behavior. First, the $\epsilon_x$ values are almost the same for different values of initial divergence for both $a_0=20$  and $a_0=40$, which means that the value is dominated by $a_0$. Second, $\epsilon_y$ increases with the increase of the initial electron beam emittance, but overall remains relatively small due to the fact that the electron beam loses a significant fraction of its energy and positrons are predominantly produced with low energy. Third, while for $a_0=20$ the number of produced positrons is about 2\% of the initial electron number, at $a_0=40$  the number of produced positrons  goes up to 30\%. Thus, an electron beam with a 100 nm emittance focused down to 100 nm transversely and having 9.1 GeV energy is able to produce a positron beam with the total charge of about 30\% of the initial electron beam charge and with 300 nm emittance. (Here we assumed that the emittance exchange can be achieved for initially $\epsilon_x=0.62$ $\mu$m and $\epsilon_y=0.06$ $\mu$m  positron beam.) This is approaching collider relevant emittances, as Ref.~\cite{benedetti.arxiv.2022b} gives $\epsilon_n=100$ nm as a desirable starting point for the electron and positron sources.

In principle, further reduction of the positron beam emittance can be achieved by employing laser pulses with higher frequencies. If the laser energy is fixed, than increasing the frequency leads to the reduction of $a_0$. Lower $a_0$ results in lower emittance, but with frequency increase we maintain the same value of the field, and, thus, the same value of parameters $\chi_e$ and $\chi_\gamma$, which determine the rate of the MC and MBW processes. For example, if we take $a_0=20$ and $\omega=2\omega_0$, then the resulting emittance of the positron beam is $\epsilon_x=0.29$ $\mu$m and $\epsilon_y=0.02$ $\mu$m, where $\epsilon_x$ is two times smaller than in the $a_0=40$, $\omega=\omega_0$ case. Further increase of the laser frequency, $\omega=4\omega_0$, leads to further reduction of the positron beam emittance as shown in Table \ref{tab:4omega}, where the same parameters of the electron beam and the same laser field strengths as in Table \ref{tab:1omega} are used, down to $\epsilon_x=0.13$ $\mu$m and $\epsilon_y=0.02$ $\mu$m for different initial electron beam emittances. Thus, frequency upshift for the colliding laser pulse opens a path to collider-relevant sources of positrons. 

\begin{table}[]
\begin{tabular}{|c||cc|cc|}
\hline
\multirow{2}{*}{$\epsilon_x/\epsilon_y$} & \multicolumn{2}{c|}{$a_0=5,~\omega=4\omega_0$}    & \multicolumn{2}{c|}{$a_0=10,~\omega=4\omega_0$}    \\ \cline{2-5} 
                  & \multicolumn{1}{l|}{electron} & positron & \multicolumn{1}{l|}{electron} & positron \\ \hline \hline
 $\theta=0.1,~\epsilon_0=0.1$     & \multicolumn{1}{l|}{0.06/0.04} & 0.08/0.026 & \multicolumn{1}{l|}{0.1/0.02} & 0.13/0.02 \\ \hline
 $\theta=0.25,~\epsilon_0=0.2$    & \multicolumn{1}{l|}{0.11/0.096} & 0.095/0.05 & \multicolumn{1}{l|}{0.1/0.016} & 0.13/0.016 \\ \hline
 $\theta=0.5,~\epsilon_0=0.48$     & \multicolumn{1}{l|}{0.2/0.2} & 0.13/0.1 & \multicolumn{1}{l|}{0.1/0.016} & 0.13/0.016 \\ \hline
\end{tabular}
\caption{The normalized emittances, $\epsilon_x$ and $\epsilon_y$ of electron and positron beams after the interaction with either $a_0=5$ or $a_0=10$ and upshifted frequency $\omega=4\omega_0$ laser pulse for three initial values of electron beam divergence $\theta=0.1,~0.25,~0.5$ mrad. For $a_0=10$, all three cases of initial divergence result in the production of a 40 pC positron beam. For $a_0=5$ the positron beam charge is down to 2.4 pC.}
    \label{tab:4omega}
\end{table}

The divergence of the photon beam generated in the interaction of the electron beam with the linearly polarized laser pulse, with $a_0\ll 1$, is $1/\gamma$ along the axis orthogonal to the laser polarization and $a_0/\gamma$ along the laser polarization axis \cite{gonoskov.rmp.2022, blackburn.pop.2018, blackburn.pra.2020}. However, the positrons clearly do not follow this distribution. It is due to the fact that  positrons are charged particles, thus, their motion is affected by the EM fields of the laser pulse. Moreover, the positron ejection angle from the laser pulse  is determined by its initial energy and the phase in the laser pulse  where it was produced \cite{heinzl.pre.2015}. Since the BW process has a strong dependence on the field strength, positrons are mainly produced near the maximum of the field, which is simultaneously the minimum of $a_0$. Thus, the field structure of the linearly polarized laser pulse leads to reduced emittance of a positron beam, cf. Tables \ref{tab:1omega} and \ref{tab:4omega}.    

\section{Conclusions and outlook.}
\label{sec:conclusions}

Laser plasma based acceleration of charged particles is considered a promising concept for a number of potential applications ranging from sources of high frequency radiation for imaging and national security to future lepton and $\gamma\gamma$ colliders for fundamental physics studies. While many of these applications need just high energy electrons, some of them require positron beams. The laser-plasma-based acceleration of positrons still remains a challenge, partly due to the lack of positron sources with properties tailored for LPA applications. In particular, it is desirable to avoid the utilization of damping rings to accumulate and cool down positron beams. Thus, a positron source able to produce beams with substantial charge is needed. Recently several schemes of positron production for suitable for LPA applications were proposed and studied (see Refs. \cite{amorim.ppcf.2023, terzani.arxiv.2023} and references cited therein), however, these sources tend to produce beams with large emittance and angular divergence. 

Here we studied a positron source based on a high-energy electron beam interaction with a high-intensity laser pulse. We assumed that a laser pulse can be split into two beams: one to accelerate electrons via LPA, and the other to be tightly focused to provide an high intensity EM field. These accelerated electrons are to collide with this high-intensity field at the interaction point. During such interactions, electrons emit high-energy photons (multiphoton Compton process), which in turn decay into electron-positron pairs (multiphoton Breit-Wheeler process), giving rise to a positron beam. It was found that a PW-class laser can be configured to provide a setup featuring a multi-GeV electron beam interaction with a $\sim 10^{21-22}$ W/cm$^2$ laser pulse. As a result of this interaction positron beams with a charge of tens to hundreds of pC can be generated with a duration that is of the order of the initial electron beam duration. Moreover, these beams demonstrate low values of angular divergence and emittance, which are crucial prerequisites for collider applications.  

Since this positron source is based on the consecutive MC and MBW processes, which strongly depend on the particle energy and EM field strength, the source can be optimized from the point of view of the positron beam charge and emittance. We found that while the total laser energy used to accelerate electrons and provide strong EM field is an important input parameter, the radius of the initial electron beam is more crucial, in terms of high charge and low emittance positron beam production. The reduction of the electron beam radius from being equal to that of the laser focal spot to 100 nm increases the number of produced positrons by an order of magnitude. So that 40 J of total laser energy is able to generate a 40 pC positron beam. Also, tightly focused electron beams have low emittance, which is imprinted on the positron beams. Moreover, the strong dependence of the MBW process on the field strength further reduces the emittance of the positron beams, owing to the production of electron-positron pairs near the maximum of the EM field, which is also the minimum of the vector potential in the case of linearly polarized field. The final transverse momentum of positrons is proportional to the value of the vector potential at the EM field phase where the electron-positron pair was born via the MBW process. 

The simulations of a 9.1 GeV, 100 nm radius, electron beam interaction with $a_0=20$ and $a_0=40$ laser pulses show the production of positron beams with $\epsilon_x=300$ nm, $\epsilon_y=50$~nm and $\epsilon_x=600$ nm, $\epsilon_y=30$ nm normalized emittances, respectively. While $\epsilon_x$ values are dominated by the value of the vector potential, $\epsilon_y$ values are determined by the angular dependence of the MBW process. Thus, the reduction of positron beam emittance can be achieved through the reduction of the EM field vector potential value in the phase where the MBW process occurs. We propose to use high frequency laser pulses in order to reduce the value of the vector potential while keeping the EM field strength the same. We simulated several cases of upshifted frequency laser pulses: $a_0=20$ and $\omega=2\omega_0$ as well as  $a_0=10$ and $\omega=4\omega_0$ and found further decrease of positron beam emittance down to $\epsilon_x=300$ nm, $\epsilon_y=20$ nm for $\omega=2\omega_0$ and $\epsilon_x=130$ nm, $\epsilon_y=20$ nm for $\omega=4\omega_0$. Thus, higher frequency laser pulses offer a way to produce positron beams with collider-relevant emittances using the interaction of an electron beam with a high intensity laser pulse.

\section*{Acknowledgments}
This research was supported by LDRD funding from LBNL provided by the Director and the U.S. DOE Office of Science Offices of HEP and FES (through LaserNetUS) under Contract No. DE-AC02-05CH11231, and used the computational facilities at the National Energy Research Scientific Computing Center (NERSC).

\section*{Appendix: Idealized LPA Stages in the Quasi-Linear and Bubble Regime}
\label{sec:Appendix}

In the following, we describe the details of the idealized LPA stages, which were used to estimate the electron beam energies given the available laser power. 

For the idealized stage in the quasi-linear regime, we considered an LPA driven by a super-matched to a parabolic plasma channel \cite{benedetti.pre.2015} laser pulse with $a_0=1.6$, $k_p w_b=4$, and $k_p c T_{fwhm}=2.12$ (Gaussian longitudinal profile). Here $k_p=(4\pi n_0 e^2/mc^2)^{1/2}$ is the plasma wavenumber and $w_b$ is the laser waist. The central laser wavelength is 800 nm. The operational density depends on the laser energy and is given by $n_0[\hbox{cm}^{-3}]\simeq 7.14\times 10^{17} (U_1 [J])^{-2/3}$. 

For the stage operating in the bubble regime, laser driver is bi-Gaussian and its intensity is such that $a_0=4.5$, furthermore laser focal spot size $w_{\mathrm{b}}$ and pulse length $\tau$ are chosen according to the theory in Ref.~\cite{lu.prstab.2007} (i.e., $k_p w_0=2\sqrt{a_0}$, and $cT_{fwhm}=(2/3)w_0$), and the central laser wavelength is 800 nm. The operational density of the stage is specified once the laser energy is specified and is given in the bubble regime as $n_0[\hbox{cm}^{-3}]\simeq 7.02\times 10^{18} (U_1 [J])^{-2/3}$. Note that, for a given laser energy, and for the parameters considered here, the density of a stage operating in the quasi-linear regime is about an order of magnitude lower compared to the one of a stage operating in the bubble regime. Due to the longer dephasing and depletion lengths at lower densities, the energy gain provided by a quasi-linear stage is generally larger than that provided by a stage operating in the bubble regime.

In both, the quasi-linear and bubble stages, the initial electron beam is chosen to experience $\sim$75\% of the maximum accelerating field (for the stage in the bubble regime the maximum field is obtained with a linear extrapolation of the longitudinal wake to the back of the bubble), and the current profile is such that the longitudinal wakefield in the beam region is initially   flat (i.e., strongly beamloaded stages). The charge of the electron beam is $Q_b[\hbox{pC}]\simeq 37 (U_1 [J])^{1/3}$ in the quasi-linear stage, and $Q_b[\hbox{pC}]\simeq 139 (U_1 [J])^{1/3}$ for the bubble case.

\bibliography{positrons.bib}

\end{document}